\documentstyle[aps,epsfig,multicol]{revtex}
\begin{document}
\draft

\title{
Metastable States of a Coupled Pair on a Repulsive Barrier
}
\author{F.M.Pen'kov}
\address{Joint Institute for Nuclear Research,\\
141980, Dubna, Russia\\
penkov@thsun1.jinr.ru}
\date{\today}
\maketitle

\begin{abstract}
Resonance penetration of two coupled particles through
a repulsive barrier is considered.
It is shown that a local minimum of the total potential generates
metastable bound states, and their spectrum determines
the position of resonances in the penetration probability.
It is pointed out that the probabilities of tunneling of
two interacting particles from the false vacuum can be
essentially higher than it has been assumed earlier.
\end{abstract}

\begin{multicols}{2}

In paper by N.Saito and Y.Kayanuma~\cite{Saito}, it was pointed
out that there exists a new quantum phenomenon -- resonance
transparency of a single repulsive barrier for a coupled pair of
particles.
To consider this effect, a one-dimensional rectangular repulsive
barrier and an infinite one-dimensional rectangular potential
well coupling the pair were chosen.
Since the interactions were simple, it was possible to solve the
initial two-dimensional Schroedinger equation by reducing it to
a system of one-dimensional equations by means of projection onto
7 eigenfunctions.
However, there still remained the question of how this effect
manifests itself in other systems.

This note continues the study of the effect of resonance
transparency of two type one-dimensional barriers for a pair of
identical particles coupled by the oscillatory interaction. This
pair interaction allows us to reduce the problem of
three-dimensional scattering of a three-dimensional oscillator to
the solution of a two-dimensional equation analogous to the
equation derived in ref.~\cite{Saito}.
Besides, it is just this sort of pair interaction that is used in
literature~\cite{Rub1} devoted to the probability of induced
decay of the false vacuum in collisions of high-energy particles
(see, for instance,~\cite{Rub2,Tyn1}).
It was pointed out there that it is possible to describe the
processes of transition from the false vacuum on the basis of
quantum-mechanical tunneling of a pair of particles through
the barrier; but the study was performed for a system where only
one of the oscillator particles interacts with the barrier.
Here, we show that, when the two particles interact with the
barrier, there arises the same effect of resonance transparency
as in ref.~\cite{Saito}.

The first potential barrier we study is taken in the Gauss form
from ref.~\cite{Rub1} in order to show that it is possible to
drastically increase the probability of induced decay of the
false vacuum.
The second potential barrier of the Coulomb form is investigated
in order to draw attention to the fact that the resonance
tunneling of the barrier is feasible in the problems of fusion
of heavy nuclei.
The method of investigation is based on the numerical solution
of the two-dimensional Schroedinger equation without any further
simplifications.

Consider the penetration of a pair of identical particles with
masses $m_1=m_2=m$, and coordinates
$ {\bf r}_1$ and $ {\bf r_2}$
coupled by an oscillatory interaction through the
potential barrier $ V_0(x_1)+V_0(x_2)$.
The Hamiltonian of this system ($\hbar $ = 1)
$$ -\frac{1}{4m}\triangle_R -\frac{1}{m}\triangle_r +
\frac{m\omega^2}{4} r^2 +  V_0({\bf R-r/2}) + V_0({\bf R+r/2}),
$$
written in coordinates of the center of inertia of the pair
 ${\bf R} = ({\bf r}_1 + {\bf r}_2)/2 $
and in an internal coordinate of the relative motion
 $ {\bf r}= {\bf r}_1 -{\bf r}_2 $
describes the three-dimensional motion of a three-dimensional
oscillator with the frequency of vibrations $\omega$.
Since the potential barrier depends only on one variable,
and the oscillatory interaction is additive in  ${\bf r}$
projections, the wave function is factorized, and its nontrivial
part describing the process of scattering depends only on two
variables.
It is convenient to represent these variables in the dimensionless form
 $$
 x= \sqrt{ \frac{m \omega}{2}}(x_1-x_2),
 \ \ \ \ y=\sqrt{ \frac{m \omega}{2}}(x_1+x_2).
 $$
The Schroedinger equation in these variables is of the form
 \begin{equation}
 \left( -\partial_{x}^{2} - \partial_{y}^{2} + x^2 +
 V(x-y)+V(x+y) -E \right)\Psi =0,
 \label{EqS}
 \end{equation}
where the energy $E$ is written in units  $\omega/2$, and the
potential barrier
$ V(x \pm y) = \frac{2}{\omega}V_0(( x \pm y)/\sqrt{2m\omega})$
in what follows will be written in a form convenient for us.
Equation (\ref{EqS}) should be supplemented with boundary
conditions.
Let the process of scattering proceed from left to right,
and the initial state of the oscillator be the state $n$.
Then the boundary conditions are written in the form
 \begin{equation}
 \begin{array}{lr}
   \Psi \bigl|_{y \to -\infty}= &
   \exp(i k_n y)\varphi_n(x) -
   \sum \limits_{j \leq N} S_{nj} \exp(-i k_j y) \varphi_j(x);\\
   \Psi\bigl|_{y \to + \infty}= &
   \sum \limits_{j \leq N} R_{nj} \exp(i k_j y) \varphi_j(x); \\
   \Psi\bigl|_{x \to \pm \infty}= &  0 .
  \end{array}
 \label{BC}
 \end{equation}
The wave functions of the oscillator $ \varphi_j(x)$ obey the
Schroedinger equation
\begin{equation}
\left( -\partial_{x}^2 + x^2 - \varepsilon_i \right)\varphi_i =0
\label{Osc}
\end{equation}
with the energy $ \varepsilon_j = 2 j + 1 $ ($j=0,1,2,...$),
momenta  $ k_j=\sqrt{E-\varepsilon_j}$,
and the number $N$ of the last open channel
($ E-\varepsilon_{N+1} < 0 $).
In what follows, we consider an oscillator composed of
bosons, whose spectrum is convenient to number from 1.
 So, hereafter, $ \varepsilon_j = 4 j -3 $ ($j=1,2,...$).

The probabilities of penetration  $W_{ij}$ and reflection
$D_{ij}$ are defined as the ratio of the density of
a penetrated or a reflected flux to the density of
the flux of incident particles:
$$
W_{ij}= |R_{ij}|^2 \frac{k_j}{k_i}, \ \  D_{ij}= |S_{ij}|^2
 \frac{k_j}{k_i}.
$$
It is obvious that
$ \sum_{j \leq N} \left( W_{ij}+D_{ij} \right) =1 $.

To determine the probabilities of penetration (reflection) in the
above way, it is necessary to solve the two-dimensional
differential equation (\ref{EqS});
its numerical solutions will be given below.
The numerical solution was based on a three-diagonal
approximation of second derivatives with constant steps in
 $x$ and $y$: $h_x=0.025$, $h_y=0.005$, respectively.
Finite dimensions  $|y_{\rm max}|=12$ and  $|x_{\rm max}|=7$
of the range of numerical calculations, at a given degree
of discretization, provided accuracy to within the third decimal
point in all the presented calculations.

\begin{figure}[ht]
\begin{center}
\parbox{8cm}{
\mbox{\epsfig{file=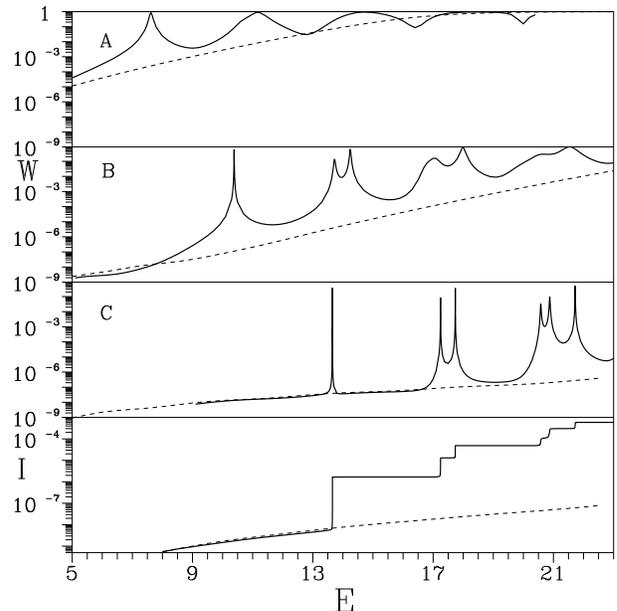,width=8cm}}
\caption {Total probabilities of penetration through the barrier.
The solid curve is for potential (\ref{potRub}); the dotted one,
for potential $2V(y)$.
The explanation is given in the text.
}
\label{FigRl}
}
\end{center}
\end{figure}

As it was indicated above, in papers devoted to the induced decay
of false vacuum, use was made of the model of quantum-mechanical
tunnelling of a pair coupled by the oscillatory interaction
through a barrier~\cite{Rub1}.
The case when only one particle interacts with the barrier was
considered.
In the framework of the resonance tunnelling, we can expect
that the picture of tunnelling would essentially change when the
interaction of both the incident particles with the barrier is
switched on.
The barrier used in ref.~\cite{Rub1}, upon being made
dimensionless, is of the form
\begin{equation}
V(X)= \frac{2}{g^2 \omega} \exp(-g^2 X^2/\omega); \ \ X=x \pm y,
\label{potRub}
\end{equation}
where $\omega$ is the oscillator frequency, and  $g^2 \ll 1$
is the model parameter of false vacuum.
In ref.~\cite{Rub1}, the dependence of the tunneling probability
was calculated in the range of $g^2$ from 0.09 till 0.01 and at
fixed frequency  $\omega=1/2$.
In the present calculations, the same value of $\omega$ but
greater values of  $g^2$ are accepted.
The reason is that there arise extremely (in the framework of
numerical calculations) narrow resonances in the energy
dependence of the tunneling probability of a coupled pair
through the barrier.
Therefore, in Fig.\ref{FigRl}, we present the results
of numerical calculations of equation(\ref{EqS})
with barrier (\ref{potRub}) at $g^2=$ 0.5, 0.3, 0.2
denoted by letters A, B, C, respectively.
%

In this Fig., we draw the probabilities of penetration of
a coupled pair from the ground state into all the possible
states, i.e. $W=  \sum_{j \leq N} W_{1j}$  at different values of
$g^2$.
A prominent resonance dependence of the tunneling probability
shows that there do exist barrier resonances under discussion.
Note that the first resonance with decreasing  $g^2$ shifts
towards higher energies, but the quantity   $g^2 E_r$
diminishes.

When $g^2 = 0.2$, the probability of resonance penetration is
many times ($\sim 10^8$) as large as the probability of
penetration in the nonresonance region (background).
Therefore, in Fig.\ref{FigRl}, we present the results of
calculations on the logarithmic scale.
They demonstrate not only the indicated exceeding but also the
coincidence of the background part of the curve with the
probability of penetration of a structureless particle,
i.e. with the solution of equation (\ref{EqS}) for
the barrier potential $2 V(y)$.
To estimate the contribution of narrow resonances to
the probability of penetration of the particle flux distributed
over energy, in Fig.\ref{FigRl}, we plot the penetrated flux
$$
I(E)= \frac{1}{E-E_{0}}  \int^{E}_{E_0} W(E') dE',
$$
in the case when the incident flux is distributed uniformly from
$E_0$ till $E$; the quantity $E_0 = 5$.
It is seen that the main contribution to the probability
of penetration comes from resonances.
At $E = 23$, the difference from the
background penetration amounts to 4 orders.

Now, we describe a simple scheme of arising barrier metastable
states that make the barrier transparent.
It is not difficult to verify that the
potential energy $U(x,y)=V(x+y) + V(x-y)+x^2$
possesses a local minimum at  $y=0$ (the center of mass in
the middle of the barrier) and at some values of $x= \pm x_0$.
A maximum is at  $x=0$.

This, there exist two potential "wells" separated by the barrier.
Bound states of that system split into even and odd states.
The magnitude of splitting is determined by the probability of
penetration through the internal barrier.
When $2V(x=0,y=0) \gg 1 $, this shift can be very small, and
the spectrum of even states is determined by the spectrum of
an isolated "well".
In the first approximation, the position of resonances can be
described by the oscillator spectrum of bound states at
$y=0$ and $x=x_0$
\begin{equation}
E_{n_x n_y}=  E_0 + 2 \omega_x (1/2 +n_x) +
2 \omega_y (1/2 +n_y),
\label{spectr}
\end{equation}
where $n_x$ and $n_y$ are oscillator quantum numbers;
$E_0 = 2 V(x_0)$; and frequencies $\omega_x$ and $\omega_y$
are determined by the second derivatives at the point
of local minimum:
$
\omega_x = \sqrt{\partial^2_x U(x,y)/2} \
$,
$
\omega_y=  \sqrt{ \partial^2_y U(x,y)/2}
$.

For potential (\ref{potRub}), it is not difficult to obtain the
oscillator-model parameters
\begin{equation}
\begin{array}{c}
E_{0} = \omega (1+2 \ln(2/\omega))/g^2;\\
[2mm]
\omega_x^2 = 4\ln(2/\omega);\\
[2mm]
\omega_y^2 = \omega_x^2 -1.
\end{array}
\label{spectrR}
\end{equation}

Let us compare the positions of resonances drawn in
Fig.\ref{FigRl} at  $g^2=0.2$ with the results of calculation
by formulae (\ref{spectr}) and (\ref{spectrR}) from which it
is clear that there exist small but clearly seen satellite
resonances.
The position of the first resonance $E_r=13.65$ is well described
by the oscillator energy in the ground state  $E_{00}=13.92$.
The second group of resonances is generated by a single
excitation of oscillators either along $y$ or along $x$:
$E_{01}=18.18$, $E_{10}=18.63$.
They are associated with the resonances at energies
$17.24$ and $17.74$.
The third group of resonances is generated by a double excitation
$E_{02}=22.45$, $E_{11}=22.90$, $E_{20}=23.34$, and respectively,
resonance energies are $20.58$, $20.88$, $21.72$.
So, the simple oscillator model of a metastable
barrier state gives a correct qualitative picture of the origin
of resonances.
Comparison with Fig.\ref{FigRl} shows that the largest values
of the tunneling probability correspond to metastable states
with a minimal excitation along the coordinate of the center
of inertia.

Decomposition around the point of equilibrium does not exhaust
all possibilities of the oscillator model.
Agreement between the resonance energy and the energy of
a metastable state can be improved by a simple
variational procedure.
To this end, we consider the position of a minimum $x_0$
and frequencies $\omega_x $ and $\omega_y$  to be unknown
quantities that are determined by the minimum of the average
of the total Hamiltonian
$$
\overline H  = \langle \phi_x \phi_y|
  -\partial_{x}^{2} - \partial_{y}^{2} + x^2 +
  V(x-y)+V(x+y)| \phi_y \phi_x \rangle
$$
over normalized eigenfunctions of the oscillator in the ground
state:
$$
\phi_x = \left( \frac{\omega_x}{2\pi}\right)^{1/4}
\exp \left(-\frac{\omega_x}{4} (x-x_0)^2 \right);
$$
$$
\phi_y = \left( \frac{\omega_y}{2\pi}\right)^{1/4}
\exp \left(-\frac{\omega_y}{4} y^2 \right).
$$
Varying $ \overline H $ over $x_0$, $\omega_x$ and $\omega_y$,
we can derive a system of three nonlinear equations
to be not presented here in view of their being cumbersome.
The of the three equations are solved analytically:
$x_0=x_0(\omega_x,\omega_y)$, $\omega_y =\sqrt{\omega_x^2-1}$.
In this way, the function of one variable
$\overline H =\overline H(\omega_y)$ is to be estimated
numerically; its minimum determines the
variational estimate $E_{\rm var}$ for the first resonance.
Note that the variational connection of $\omega_x$ and $\omega_y$
is the same as in the case of decomposition  (\ref{spectrR}).
In Table~\ref{Tvar}, we present the comparison of $E_{\rm var}$
with positions of the first resonance $E_{r} $ drawn in
Fig.\ref{FigRl}. Agreement can be considered good in the
framework of the above-indicated accuracy.
\parbox{7cm}
{
\begin{table}
\caption{ Comparison of positions of the first resonance with the
variational estimate}
\begin{tabular}{|c|c|c|c|}
$g^2$   &  $E_{\rm var}$ & $E_{r}$ & $g^2 E_r \omega/2$ \\
\hline
0.5   &  7.649  &   7.62 & 1.30  \\
0.3   & 10.416  &  10.38 & 0.779 \\
0.2   & 13.680  &  13.65 & 0.683
\end{tabular}
\label{Tvar}
\end{table}
}
When  $g^2 \ll 1$, the variational expressions get simplified and
allow the following decomposition:
$$
E_{\rm var}^{\rm as} \sim E_{00} +O(g^2).
$$
It coincides, with an accuracy up to $O(g^2)$, with the energy
derived by a simple decomposition around the minimum of $U(x,y)$.
So, the estimate of the resonance spectrum made by formulae
(\ref{spectr}) and (\ref{spectrR}) is asymptotic as $g^2 \to 0 $.
In particular, when $g^2 \to 0$, we can indicate the limiting
position of the first resonance in units of $g^2$,
i.e. the quantity  $g^2 E \omega /2$ used in ref.~\cite{Rub1}:
$$
g^2 E \omega/2 \to \omega^2 (1+2 \ln(2/\omega))/2.
$$
At $\omega=1/2$, this energy tends to $0.472$; this is shown in
the fourth column of Table~\ref{Tvar}.
In ref.~\cite{Rub1} where the study was made of the penetration
of a pair through the barrier, a smooth curve was obtained for
the probability of penetration in the energy interval from
$1.2$ to $2$.
From the presented calculations it follows that, if the
interaction of both particles with the potential barrier is taken
into account, this curve becomes essentially nonmonotone.

\begin{figure}[h]
\begin{center}
\parbox{8cm}{
\mbox{\epsfig{file= 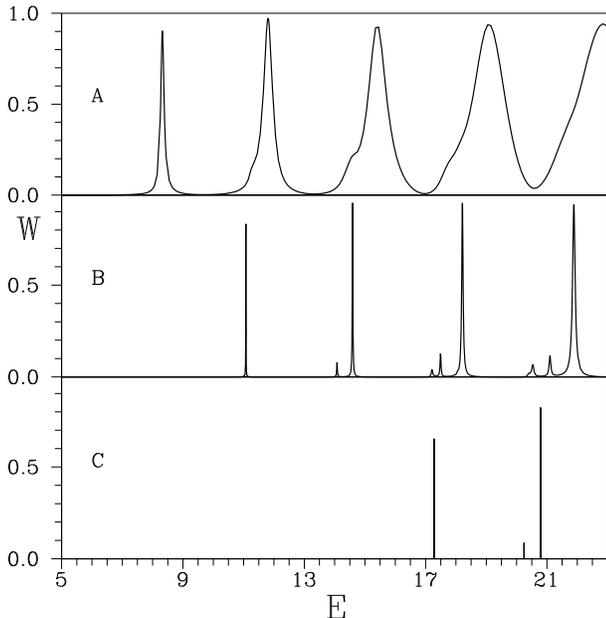,width=8cm}}
\caption {Probabilities of penetration through barriers of the
Coulomb type.
Explanations are given in the text}
\label{FigC}
}
\end{center}
\end{figure}

The barriers considered above are of the form of the Gauss
function.
For completeness, below we present the calculations for a barrier
of the Coulomb shape cutoff both at short and long distances:
\begin{equation}
V(X)= \left\{
\begin{array}{cl}
 Q /X_{\rm min}  &: |X| < X_{\rm min}\\
 Q /|X| &: X_{\rm min} \leq |X| \leq X_{\rm max}\\
 Q/X_{\rm max}  &: |X| > X_{\rm max}
 \end{array}
 \right.
 ; X= x \pm y.
\label{potCul}
\end{equation}
The cutoff at short distances was introduced for modelling
the nuclear Coulomb barrier in the framework of constraints
imposed by the one-dimensional scattering.
With this cutoff, the notion of "barrier height" is
meaningful for the one-dimensional model of scattering.
For a greater analogy, the barrier width at $|X|= X_{\rm min}$
should be small in spatial units of the problem, i.e.,
as compared with the mean-square dimension of the oscillator.
The cutoff at long distances was introduced to make it possible
to use an asymptotics of the type (\ref{BC}).
The quantity $X_{\rm max}$ should be larger than 1 for
imitating the barrier of small transparency.
Here we took $X_{\rm min}=0.1$ ¨  $X_{\rm max}=5$.
The quantity  $Q$ determines the energy height of the barrier.
In Fig.\ref{FigC}, we show the results of calculations for
$Q=$ 2, 4, 10 denoted by letters A, B, and C, respectively.

In this case, a clear picture of the resonance tunneling of
a coupled pair is also observed.
We do not present the analysis of the oscillator model
for the position of resonances, because the chosen potential
is essentially of the model character.
We only mention that satellite resonances manifest themselves
clearly, and energies of principal resonances are equidistant.

The considered mechanism of the transparency of barriers for
a coupled pair manifests itself for all the potential barriers
chosen for the investigation.
As the resonance transparency was first observed~\cite{Saito}
for barriers of the rectangular shape, and the coupling in a pair
was of nonoscillator type, it could be assumed that the resonance
transparency of barriers for composite particles could
be observed for a wide class of interactions.
Therefore, the effects of quantum transparency could occur in
various fields of physics.
In particular, when the interaction of two particles with
a barrier is taken into account, the picture of induced decay of
the false vacuum changes essentially.

The author expresses his deep gratitude to A.K.Motovilov for
the fruitful idea of realization of the numerical scheme and
to Yu.M.Chuvil'skii for pointing the importance of the study
of the resonance transparency of
Coulomb barriers.

\end{multicols}
\end{document}